\documentclass[twocolumn,showpacs,aps,prl,superscriptaddress,floatfix]{revtex4}

\usepackage{color}
\definecolor{Red}{rgb}{1,0,0}
\definecolor{Blue}{rgb}{0,0,1}
\usepackage{graphicx}
\usepackage{dcolumn}
\usepackage{bm}

\begin{document}

\preprint{}

\title{Two Dimensional Kagome Correlations and Field Induced Order in the Ferromagnetic XY Pyrochlore Yb$_2$Ti$_2$O$_7$} 

\author{K.A. Ross}
\affiliation{Department of Physics and Astronomy, McMaster University,
Hamilton, Ontario, L8S 4M1, Canada}
\author{J.P.C. Ruff} 
\affiliation{Department of Physics and Astronomy, McMaster University,
Hamilton, Ontario, L8S 4M1, Canada}
\author{C.P. Adams} 
\affiliation{Department of Physics and Astronomy, McMaster University,
Hamilton, Ontario, L8S 4M1, Canada}
\affiliation{Department of Physics, St.\thinspace Francis Xavier University, Antigonish, Nova
Scotia, B2G 2W5, Canada}
\author{J.S. Gardner}
\affiliation{Indiana University, 2401 Milo B. Sampson Lane, Bloomington, Indiana 47408, USA} 
\affiliation{National Institute of Standards and Technology, 100 Bureau Drive, MS 6102, Gaithersburg, Maryland 20899-6102, USA}
%
\author{H.A. Dabkowska} 
\affiliation{Department of Physics and Astronomy, McMaster University,
Hamilton, Ontario, L8S 4M1, Canada}

\author{Y.Qiu} 
\affiliation{National Institute of Standards and Technology, 100 Bureau Drive, MS 6102, Gaithersburg, Maryland 20899-6102, USA}
\affiliation{Department~of~Materials~Science~and~Engineering,~University~of~Maryland,~College~Park,~Maryland~20742,~USA}
%
\author{J.R.D. Copley} 
\affiliation{National Institute of Standards and Technology, 100 Bureau Drive, MS 6102, Gaithersburg, Maryland 20899-6102, USA}
\author{B.D. Gaulin} 
\affiliation{Department of Physics and Astronomy, McMaster University,
Hamilton, Ontario, L8S 4M1, Canada}
\affiliation{Canadian Institute for Advanced Research, 180 Dundas St.\ W.,Toronto, Ontario, M5G 1Z8, Canada}

\begin{abstract}

Neutron scattering measurements show the ferromagnetic XY pyrochlore Yb$_{2}$Ti$_{2}$O$_{7}$ to display strong quasi-two dimensional (2D) spin correlations at low temperature, which give way to long range order (LRO) under the application of modest magnetic fields.  Rods of scattering along $<111>$ directions due to these 2D spin correlations imply a magnetic decomposition of the cubic pyrochlore system into decoupled kagome planes.  A magnetic field of $\sim$ 0.5 T applied along the [1${\bar1}$0] direction induces a transition to a 3D LRO state characterized by long-lived, dispersive spin waves.  Our measurements map out a complex low temperature-field phase diagram for this exotic pyrochlore magnet.

\end{abstract}


\maketitle

\begin{figure}[!tb]  
\centering
\includegraphics[width=7cm]{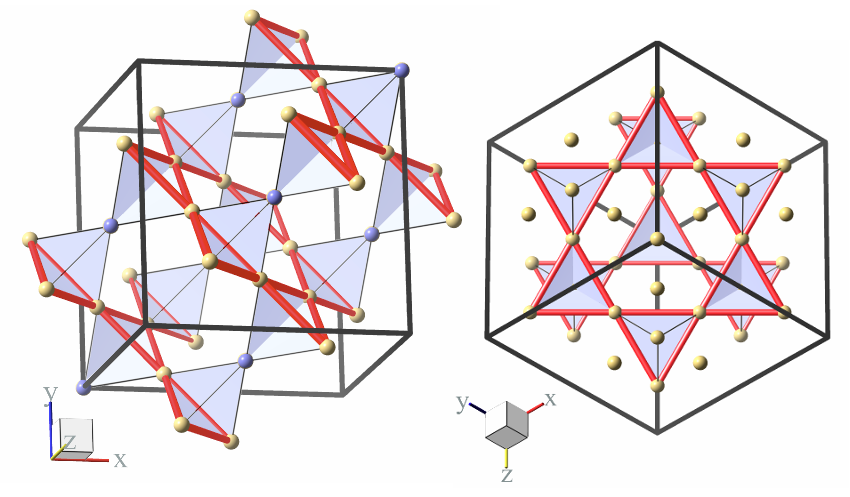}
\caption{(color online) Left: The cubic pyrochlore lattice, showing the positions of magnetic Yb$^{3+}$ ions.  This structure can be decomposed into alternating triangular (blue or dark atoms) and kagome planes (gold or light atoms), along the $<111> $ directions.  The bonds within the kagome planes are shown in red (dark) lines. Right: a view along $[111]$ shows the structure of the kagome planes.}
\label{fig:pyro}
\end{figure}

Magnetic materials displaying the pyrochlore structure are of intense current interest due to their varied and often unconventional magnetic ground states.  In such materials the magnetic ions form an array of corner-sharing tetrahedra referred to as the pyrochlore lattice, which is illustrated in Fig.\thinspace \ref{fig:pyro}.  This lattice is the 3D archetype for geometric frustration; magnetic moments residing on it cannot simultaneously satisfy all local interactions due to geometrical constraints, leading to a macroscopic degeneracy of the ground state \cite{diep}.  This frustration underlies a range of exotic disordered magnetic states at low temperatures, such as spin liquid, spin ice, and spin glass states \cite{gardner_spinliq, gardner_diffuse, bramwellSpinIce, bruce_glass}.  Many rare-earth-titanates with chemical composition $R_{2}$Ti$_{2}$O$_{7}$ crystalize as pyrochlores with the cubic space group Fd$\bar3$m.  The diverse magnetic behavior of these isostructural insulators arises from an interplay between the moment size, exchange interactions, and single-ion anisotropy resulting from crystalline electric field effects.
These often-competing interactions make these materials fertile ground for the generation of novel magnetic ground states.

Several model pyrochlore magnets have been well studied theoretically, and such studies are important benchmarks for understanding the nature of the exotic low temperature magnetic states.  For example, isotropic Heisenberg spins interacting with simple near-neighbor antiferromagnetic interactions on the pyrochlore lattice possess a disordered spin liquid ground state \cite{moessner_spinliq, canals_spinliq}.  Near-neighbour ferromagnetically-coupled spins with hard, local [111] anisotropy, such that magnetic moments point either directly into or out of the individual tetrahedra, display the disordered spin ice state at low temperatures~\cite{bram_ging}.  

In Yb$_2$Ti$_2$O$_7$, the effective spin-1/2 Yb$^{3+}$ moments exhibit overall ferromagnetic (FM) interactions, with a $\theta_{CW}$ of between +0.49K and +0.75K \cite{bramwell_mag, hodgescrysfield}, as well as easy plane anisotropy \textit{perpendicular} to the local [111] direction \cite{hodgesfluc}.  Yb$_2$Ti$_2$O$_7$ is the only known example of a ferromagnetic XY pyrochlore, and therefore is an important experimental benchmark in the field.  Intriguingly, one is not immediately lead to suspect that geometric frustration is central to the low temperature magnetic behavior of this material.  Unlike the ferromagnetic Ising-like case mentioned above, the ferromagnetic XY pyrochlore is expected to have a {\it non-degenerate} classical ground state.  Nevertheless, in this Letter we show that a disordered state is maintained in Yb$_2$Ti$_2$O$_7$ to temperatures well below both $\theta_{CW}$ and the nominal T$_C$ for this system.  We further demonstrate the ease with which a small perturbation, specifically the application of a [1${\bar1}$0] magnetic field, can drive this system to stable long range magnetic order (LRO).  Both of these phenomena are hallmarks of geometrical frustration.  Emergent frustration in a quantum system not expected to be frustrated classically has previously been discussed in relation to other rare-earth-metal titanates \cite{molavian}.  Indeed, the spin-liquid state of Tb$_2$Ti$_2$O$_7$ is known to resist LRO to temperatures at least three orders of magnitude lower than the characteristic Curie-Weiss constant \cite{gardner_spinliq}, despite being approximated by a classical $<111>$ Ising antiferromagnet model with a unique ground state \cite{enjalran}.

\begin{figure}[!tb]  
\centering
\includegraphics[width = 6.5 cm]{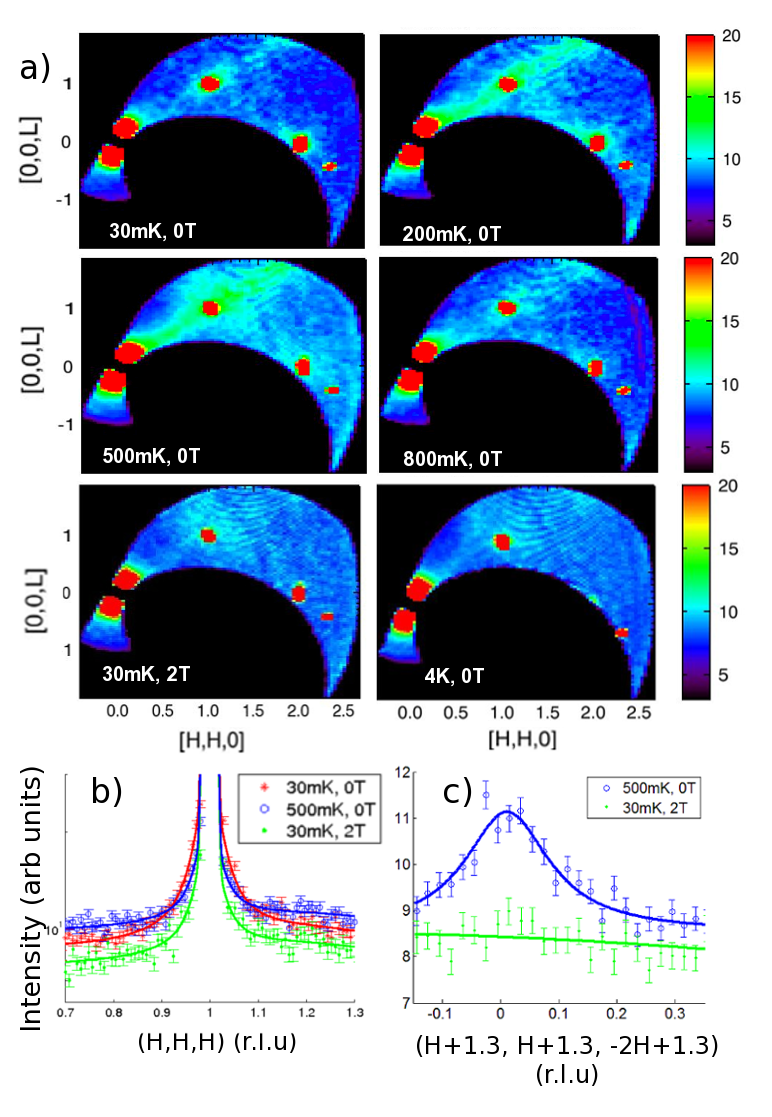}
\caption{(color online)a) Color contour maps of elastic neutron scattering in the HHL plane.  In zero applied field, diffuse rods of scattering that extend along the [111] direction are visible at 800mK and below.  The diffuse scattering in applied field (30mK, 2T) is similar to the disordered state (4K, 0T). b) Cuts of the scattering along the rod direction show the evolution of this 2D scattering with temperature and field.  c) Cuts perpendicular to the rod direction at 500mK, 0T, and within the field-induced LRO state at 30mK, 2T.  
}
\label{fig:rods}
\end{figure}

A previous study of Yb$_2$Ti$_2$O$_7$ has revealed a sharp specific heat anomaly near T$_{C}\sim240$mK \cite{blote}.  Yasui \textit{et al} presented single crystal elastic neutron scattering results that indicate simple, collinear FM ordering below T$_{C}$ \cite{yasui}.  However,  this type of ordering is inconsistent with the results of a polarized neutron scattering study \cite{gardner}.  Additionally, powder neutron diffraction measurements by  Hodges \textit{et al} indicate no obvious signs of long range order below T$_C$ \cite{hodgesfluc}.  Using M\"ossbauer and $\mu$SR techniques, this group instead argued that the nature of the transition is comparable to that of a gas-liquid system, in that the fluctuation rate of the  $\sim$3$\mu_{\text{B}}$ Yb$^{3+}$ moments drops by up to two orders of magnitude at T$_C$, but does not vanish \cite{hodgesfluc}.  Most intriguingly, a single crystal neutron diffraction measurement of Yb$_2$Ti$_2$O$_7$ revealed ``rods'' of diffuse magnetic scattering along the [111] direction of reciprocal space \cite{bonvillehyp}.  These rods indicate an unexpected 2D correlated spin state that is present both above and below T$_{C}$.   
 
\begin{figure*}[tbp] 
   
   \includegraphics[width=18cm]{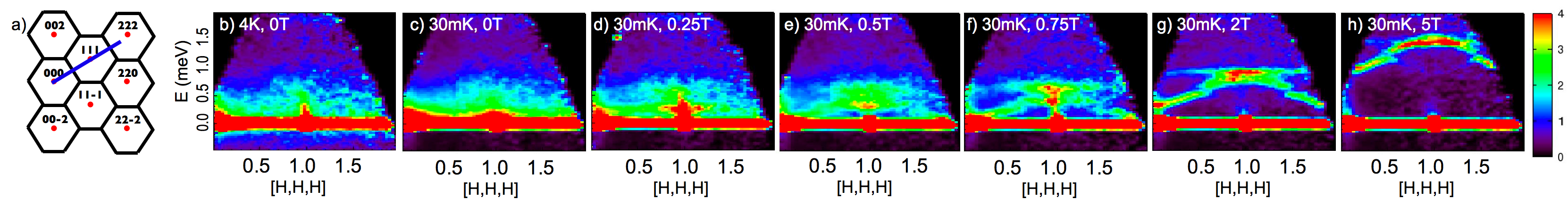} 
   \caption{(color online) Color contour maps of the inelastic neutron scattering along the [HHH] direction.  The data is binned in [H,H,-2H] over the range [-0.1,0.1].  Panel a) is a schematic of the nuclear zones in the HHL plane, and the blue line shows the direction of the cut in the inelastic panels.  Panels b) through d) show the absence of well defined spin excitations in low field, even at 30mK (panel c and d).  Resolution-limited spin waves appear at all wavevectors upon application of a magnetic field greater than $\sim$ 0.5T (panels e through h), indicating a LRO state.}
   
     \label{fig:figure3}
\end{figure*}

%


A single crystal of Yb$_2$Ti$_2$O$_7$ was prepared at McMaster University using a two-mirror floating zone image furnace.  It was grown in 4 bars of oxygen, at a rate of 5 mm/h, and the growth 
procedure closely resembled that reported previously for other rare earth titanates \cite{Gardner_growth}.  Recently, a single crystal of Y$_2$Ti$_2$O$_7$ prepared in this manner has been shown to have an exceptionally long phonon mean-free-path (10$\mu$m at 2K), indicative of the very high quality of these crystals \cite{y2ti2o7}. The $\sim$7 gram single crystal of Yb$_2$Ti$_2$O$_7$ was mounted in a vertical field dilution refrigerator cryomagnet with its (HHL) plane coincident with the horizontal plane.  Time-of-flight neutron scattering measurements were performed using the Disk Chopper Spectrometer at the NIST Center for Neutron Research\cite{DCS}.  Measurements were performed using 5 {\AA} incident neutrons, allowing us to map out the pattern of diffuse scattering in the (HHL) plane as a function of temperature and magnetic field, as well as to investigate inelastic scattering with an approximate energy resolution of $\sim$ 0.09 meV.

In zero magnetic field, elastic scattering at low temperatures within the (HHL) plane confirms the presence of rods of magnetic diffuse scattering along the [111] direction (Fig.\thinspace \ref{fig:rods}a), indicating 2D magnetic correlations in planes perpendicular to this direction.  The rods are visible at 800mK and become more clearly defined as the temperature is lowered to 200mK.  At the lowest measured temperature, 30mK, the diffuse scattering remains extended along $[111]$, but intensity has accumulated around the (111) Bragg peak (Fig.\thinspace \ref{fig:rods}b).  This indicates that 3D interplane correlations have developed at 30mK, although these are short range.

The 3D cubic pyrochlore lattice can be naturally decomposed along $<111>$ into interleaved kagome and triangular planes, and we argue that the 2D spin correlations form within the kagome planes (Fig.\thinspace \ref{fig:pyro}).  This is in view of the fact that the exchange interactions are expected to be stronger between moments within the kagome rather than the triangular planes, due to their 1st, rather than 3rd near-neighbor separation.  The question of what prevents the correlations from extending between the kagome planes is puzzling in light of the underlying cubic symmetry of the lattice, which requires that the interactions between each neighboring spin be equivalent.  Nonetheless, we conclude that only three of four spins per tetrahedron are correlated to a significant degree above $\sim$ 200mK.

This recalls the frustrated antiferromagnetic spinel Li$_2$Mn$_2$O$_4$, whose magnetic pyrochlore sublattice also decouples into kagome planes at a range of temperatures above the N\'eel temperature \cite{wiebe, wills}.  However, in the case of Li$_2$Mn$_2$O$_4$ the lattice is known to be tetragonally distorted, whereas no evidence for a departure from cubic symmetry has been reported for Yb$_2$Ti$_2$O$_7$.  We note that the diffuse rods of magnetic scattering in Yb$_2$Ti$_2$O$_7$ appear along all measured $<111>$ directions, consistent with either a spatially (domain) or temporally averaged magnetic structure.  Temporal fluctuations would require the rods be dynamic in nature, as will be shown to be the case.

Related to modifications of the rods of diffuse scattering upon cooling from 500mK to 30mK, we observe a small increase ($\sim$3\%) in the total intensity of the (111) Bragg peak, in qualitative agreement with earlier studies\cite{yasui,gardner}.  This increase does not correspond to a fully ordered state; rather, it results from the diffuse rods of scattering coalescing around Bragg peaks, an effect that is illustrated by cuts along the rods, shown in Fig.\thinspace 2b).  A similar broad feature is observed at (${\bar1}$${\bar1}$${\bar3}$), producing a $\sim$2\% increase in intensity at this peak.  The observed increase of scattering near the base of these Bragg peaks at 30mK indicates a developing short range ordered 3D magnetic state. Aside from the increase of diffuse scattering at (111) and (${\bar1}$${\bar1}$${\bar3}$), no other changes in Bragg intensities were observed upon cooling from  500mK, measured out to Q = 2.34 \AA$^{-1}$.  This is in disagreement with the increase in (222) Bragg scattering on cooling from 300mK reported by Yasui $\textit{et al }$, and we note that like the polarized neutron work \cite{gardner}, our result is inconsistent with the LRO simple collinear FM state proposed therein \cite{yasui}.  


Application of even a weak magnetic field along [1${\bar1}$0] dramatically alters the low temperature magnetic state.  Strikingly, sharp spin waves are evident at all wavevectors in the inelastic neutron scattering (INS) spectrum for magnetic fields as low as 0.75T (Fig.\thinspace \ref{fig:spinwaves}~f).  Meanwhile, the diffuse rods of scattering are eliminated.  These results demonstrate that a magnetic LRO state is induced by the application of a [1${\bar1}$0] magnetic field.  Such perturbation-induced order is also observed in other magnetic pyrochlore systems; Tb$_2$Ti$_2$O$_7$ can be driven from a spin liquid to a LRO state upon application of a [1${\bar1}$0] magnetic field \cite{rule_tbtio}, as well as as by combinations of uniaxial and hydrostatic pressure and magnetic fields \cite{mirebeau, mirebeau_nature}.


 The evolution of the inelastic scattering with the [1${\bar1}$0] applied field is shown in Fig.\thinspace \ref{fig:spinwaves}.  The spectrum is displayed along [HHH], which coincides with the direction of the rods in the low temperature, low field state.  At $\mu_0$H=0T, T=4K (panel b), diffuse, quasi-elastic (QE) scattering extends in energy to 0.6meV.  The QE scattering is most intense along the rod direction, indicating that weak, dynamic rods are present at 4K in zero field.  On cooling to 30mK, the intensity of the QE scattering increases, but its energy scale remains constant.  The corresponding fluctuation rate, $\sim$ 145 GHz, is much higher than that reported by Hodges \textit{et al} in their $\mu$SR study \cite{hodgesfluc}.  It also shows no temperature dependence in contrast to the first order-like transition they reported.  
The differences between this and the present study suggest two energy scales characterizing the spin fluctuations, which are accessible individually by the $\mu$SR and INS techniques.  The energy scale reported here is exclusive to the rods of scattering, and hence to the short range correlations.
 
 \begin{figure}[htbp] 
   \includegraphics[width = 9cm]{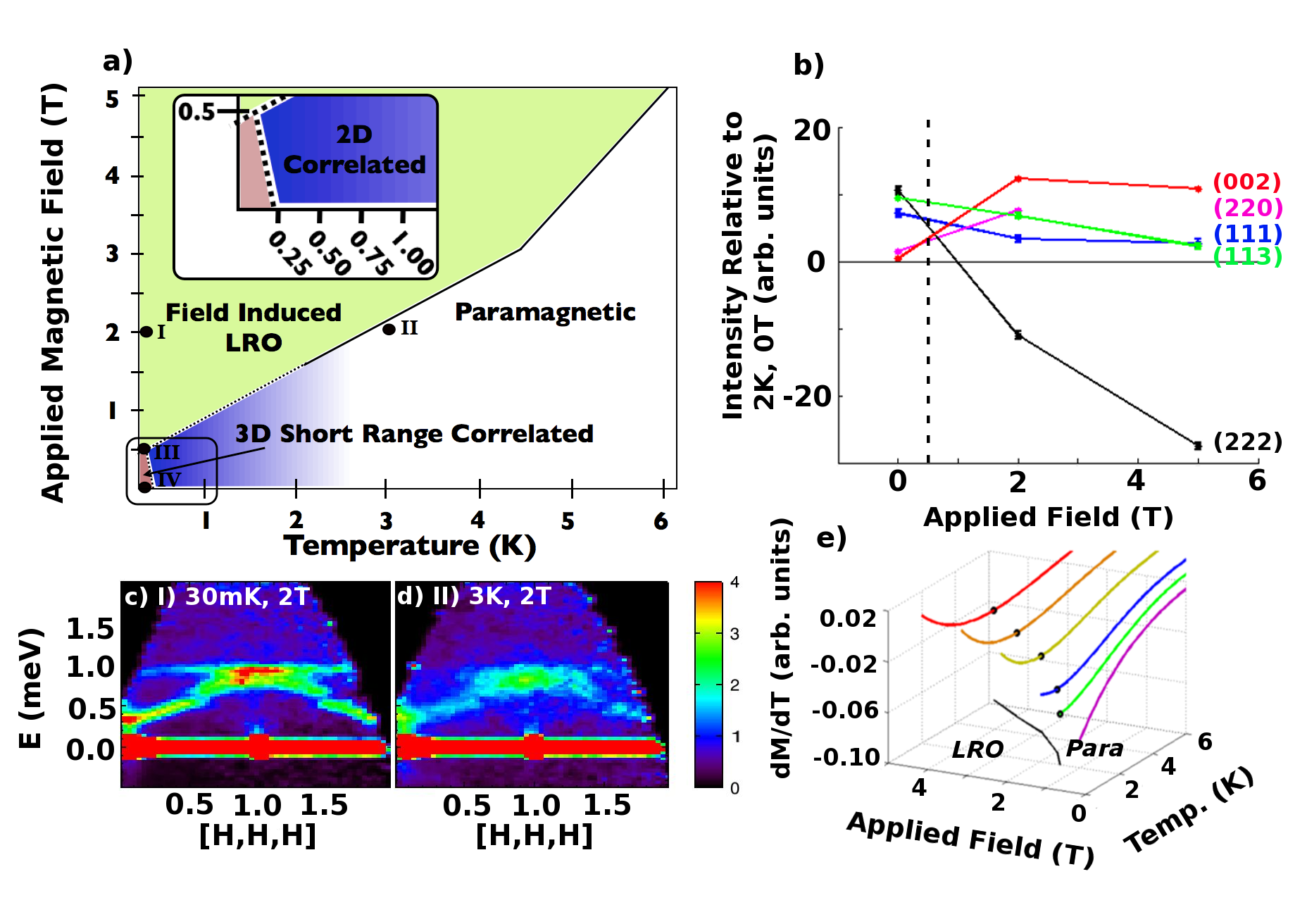} 
  \caption{(color online) a) Phase diagram for Yb$_2$Ti$_2$O$_7$.  At 30mK, the transition to LRO is indicated near 0.5T (inset), which is evinced by inelastic neutron data at points III and IV (shown in Fig.\thinspace \ref{fig:spinwaves}c and e).  b) Changes in elastic Bragg scattering as a function of field at 30mK.  c) and d) Inelastic neutron scattering data at points I and II confirm the phase boundary, which is measured from magnetization data (panel e) above T=2K.}
   \label{fig:phase_diagram}
\end{figure}

 Notably, no spin wave modes appear in the low temperature zero-field state, which is consistent with short ranged, 3D correlations, as opposed to a LRO state which would support spin wave excitations.  At $\mu_0$H=0.25T, T=30mK, the QE rod of scattering lifts off from the elastic channel, and at $\mu_0$H=0.5T a dispersion of this excitation into spin wave modes is observed (panel d and e).  In stronger applied fields, the spin waves are increasingly gapped by the Zeeman energy and become nearly resolution-limited, having a full width of approximately 0.1meV at 2T, 30mK (panel g).  Three spin wave branches are observed in this field-induced state; one with little or no dispersion while the other two exhibit minima tending to zero energy (i.e. Goldstone modes) at the (111) and (222) positions.  We argue that these dispersive modes are indicative of a LRO state on the basis that they are resolution-limited at all wavevectors, particularly at the magnetic zone centers where this quality indicates a large correlation length to the long wavelength dynamics.    The presence of two types of branches, one dispersionless and the other Goldstone-like, may be associated with fluctuations out-of and within the easy XY plane, respectively.  Indeed, a related spin wave spectrum is observed for the XY AF pyrochlore Er$_{2}$Ti$_{2}$O$_{7}$ \cite{ruff}.  In Yb$_{2}$Ti$_2$O$_7$, the magnetic zones are well-defined and coincident with the nuclear zones, as indicated by clear gapping of the dispersive branches at the nuclear zone boundaries.  We therefore identify $\mu_0$H$\sim$0.5T at T=30mK as a transition to a long ranged ordered magnetic state, and note that the coincidence of magnetic and nuclear Bragg peaks presents a challenge for the determination of the LRO structure.

Taking well-defined spin waves at all wavevectors as a defining characteristic of the field-induced LRO state, we propose the phase diagram for Yb$_2$Ti$_2$O$_7$ shown in Fig.\thinspace \ref{fig:phase_diagram}a).  To further characterize the phase diagram, magnetization measurements were performed on a small single crystal of Yb$_2$Ti$_2$O$_7$ using a SQUID magnetometer with the field aligned along [1${\bar1}$0].  
A minimum is observed in dM/dT(T), which we have mapped as a function of field (Fig.\thinspace \ref{fig:phase_diagram}e) for temperatures of 2K and above.  The resulting phase boundary is consistent with our spin wave measurements, in particular with the spin wave spectrum at points I and II, shown in Fig.\thinspace \ref{fig:phase_diagram}c) and d), which are characteristic of the LRO state and nearly paramagnetic state, respectively.  

The changes in the elastic Bragg scattering with field at T=30mK, shown in Fig.\thinspace \ref{fig:phase_diagram}b), merit comment.  Subtraction of nuclear Bragg intensities (from the 2K, 0T set) reveals that the change in (222) intensity as the field is increased past H$_{C}$, while a robust result, is anomalous since the intensity falls below the purely nuclear scattering level.  This could indicate a magnetically driven structural distortion of the lattice, which necessarily hinders the determination of the magnetic structure based on these intensity changes.  




In conclusion, Yb$_2$Ti$_2$O$_7$ exhibits spontaneous 2D correlations at low temperature, which emerge via unknown mechanisms from a seemingly isotropic cubic lattice.  The previously reported transition in fluctuation rate at $\sim$240mK involves a crossover from this 2D correlated state to a short-ranged 3D correlated state.  Application of a small magnetic field relieves the frustration and induces long range magnetic order.  Upon entering the ordered state, anomalous changes in Bragg scattering indicate the possibility of magneto-elastic coupling in Yb$_{2}$Ti$_{2}$O$_{7}$, which may ultimately underlie the emergent planar spin correlations in zero field.


The authors would like to acknowledge M.J.P.Gingras, J.P. Clancy and G. MacDougall for useful discussions, as well as P. Dube for technical assistance.  We acknowledge the use of the DAVE software package.  This work utilized facilities supported in part by the National Science Foundation under Agreement No. DMR-0454672, and was supported by NSERC of Canada.

\end{document}